\def\year{2020}\relax
\documentclass[letterpaper]{article} 
\usepackage{aaai20}  
\usepackage{times}  
\usepackage{helvet} 
\usepackage{courier}  
\usepackage[hyphens]{url}  
\usepackage{graphicx} 
\usepackage{graphicx}  
\usepackage{algorithm}
\usepackage{amsfonts}  
\usepackage{mathtools}
\usepackage{multirow}
\usepackage{subcaption}
\usepackage{balance,lineno}
\usepackage{algpseudocode}
\usepackage{balance,lineno}
\usepackage{float}
\title{Weak Supervision for Fake News Detection via \\Reinforcement Learning }
\author{Yaqing Wang,\textsuperscript{\rm 1}  Weifeng Yang,\textsuperscript{\rm 2}
Fenglong Ma,\textsuperscript{\rm 3}
Jin Xu,\textsuperscript{\rm 2}\thanks{Corresponding authors.}
Bin Zhong,\textsuperscript{\rm 2}
Qiang Deng,\textsuperscript{\rm 2}
Jing Gao\textsuperscript{\rm 1}\textsuperscript{*} \\ 
\textsuperscript{\rm 1}State University of New York at Buffalo, New York, USA \\
\textsuperscript{\rm 2}Data Quality Team, WeChat, Tencent Inc., China \\
\textsuperscript{\rm 3}Pennsylvania State University, Pennsylvania, USA  \\
\textsuperscript{\rm 1}\{yaqingwa,  jing\}@buffalo.edu, 
\textsuperscript{\rm 2}\{curryyang, jinxxu, harryzhong, calvindeng\}@tencent.com, \textsuperscript{\rm 3}fenglong@psu.edu
}
 
\begin{document}

\maketitle

\begin{abstract}
Today social media has become the primary source for news. Via social media platforms, fake news travel at unprecedented
speeds, reach global audiences and put users and communities at great risk. Therefore, it is extremely important to detect fake news as early as possible. Recently, deep learning based approaches have shown improved performance in fake news detection. However, the training of such models requires a large amount of labeled data, but manual annotation is time-consuming and expensive. Moreover, due to the dynamic nature of news, annotated samples may become outdated quickly and cannot represent the news articles on newly emerged events. Therefore, how to obtain fresh and high-quality labeled samples is the major challenge in employing deep learning models for fake news detection.  In order to tackle this challenge, we propose a reinforced weakly-supervised fake news detection framework, i.e., {WeFEND}, which can leverage users' reports as weak supervision to enlarge the amount of training data for fake news detection. The proposed framework consists of three main components: the annotator, the reinforced selector and the fake news detector. The annotator can automatically assign  weak labels for unlabeled news based on users' reports. The reinforced selector using reinforcement learning techniques chooses high-quality samples from the weakly labeled data and filters out those low-quality ones that may degrade the detector's prediction performance. The fake news detector aims to identify fake news based on the news content. We tested the proposed framework on a large collection of news articles published via WeChat official accounts and associated user reports. Extensive experiments on this dataset show that the proposed {WeFEND} model achieves the best performance compared with the state-of-the-art methods.
\end{abstract}

\section{Introduction}
The recent proliferation of social media has significantly changed the way in which people acquire information. According to the 2018 Pew Research Center survey, about two-thirds of American adults (68\%) get news on social media at least occasionally. Fake news, which refer to intentionally and verifiably false news stories, can spread virally on social media platforms as people rarely verify the source of the news when sharing a news article that sounds true. The spread of fake news may bring many negative impacts, including social panic and financial loss. Recent years have witnessed a number of high-impact fake news spread regarding terrorist plots and attacks, presidential election, and various natural disasters.
In many of these cases, even when correct information later disseminates, the rapid spread of fake news can have devastating consequences. Therefore, there is an urgent need for the development of automatic fake news detection algorithms which can detect fake news as early as possible and help stop the viral spread of such news.

Recently, many approaches are proposed to identify fake news, which can be roughly divided into two categories, i.e., traditional learning~\cite{conroy2015automatic,tacchini2017some} and deep learning based models ~\cite{ruchansky2017csi,ma2016detecting,wang2018eann,popat2018declare}. Traditional learning methods typically extract features from news articles and train classifiers based on the extracted features. Compared with traditional learning methods, deep learning models have achieved an improvement in the performance of fake news detection due to their powerful abilities of learning informative representations automatically. However, training deep learning models usually requires a large amount of hand-labeled data, i.e., news articles that are labeled as real or fake. The creation of such data is  expensive and time-consuming. Also, accurate labels can only be obtained when the annotators have sufficient knowledge about the events. Furthermore, the dynamic nature of news articles leads to decaying quality of existing labeled samples.  Some of these samples may become outdated quickly and cannot represent the news articles on newly emerged events. To maintain the quality of labeled samples, annotators have to continuously label newly emerging news articles, which is infeasible.  To fully unleash the power of deep learning models in fake news detection, it is essential to tackle the challenge of labeling fake news. 

A possible solution is to leverage the feedback provided by users who read the news. Nearly every social medial platform provides a way for users to report their comments about the news, and some of these comments are highly relevant to fake news detection. For example, for a news article published on a WeChat official account\footnote{WeChat is a Chinese multi-purpose messaging, social media and mobile payment app developed by Tencent. Wechat official accounts push news articles and information for subscribed followers.}, a user who reads the article can report whether this news is fake or not with a brief explanation. Such reports from users can be regarded as ``weak'' annotation for the task of fake news detection. The large collection of user reports can help alleviate the label shortage problem in fake news detection. However, different from expert-labeled samples, these weak annotated samples are unavoidably noisy. Users may report real news as fake ones, and the reasons they provide may not be meaningful. Therefore, how to transform weak annotation to labeled samples in the training set and select high-quality samples is the major issue we need to solve. 

In light of the aforementioned challenges, we propose a reinforced WEakly-supervised FakE News Detection framework~({WeFEND}), which can leverage the crowd users' feedback as  weak supervision for fake news detection. The proposed framework {WeFEND} consists of three main components: the annotator, the fake news detector and the reinforced selector. In particular, given a small set of labeled fake news samples together with users' feedback towards these news articles, we can train an annotator based on the feedback, which can then be used to automatically assign  weak labels for those unlabeled news articles simply based on the user feedback they received. The reinforced selector which employs reinforcement learning techniques then selects high-quality samples from weakly labeled samples as the input to the fake news detector. The fake news detector finally assigns a label for each input article based on its content. The three components integrate nicely and their interactions mutually enhance their performance. We conduct extensive experiments on a large collection of news articles published by WeChat official accounts and corresponding feedback reported by users on these articles. Experimental results show that the proposed framework {WeFEND} outperforms the state-of-the-art approaches on fake news detection. Moreover, we will publicly release this dataset\footnote{\url{https://github.com/yaqingwang/WeFEND-AAAI20}} to the community to encourage further research on fake news detection with user reports. 

The main contributions of this paper can be summarized as follows:
\begin{itemize}
\item We recognize the label shortage issue and propose to leverage user reports as weak supervision for fake news detection on news content. Towards this end, we propose an effective weakly-supervised fake news detection framework.

\item The proposed {WeFEND} framework can  automatically annotate news articles, which help enlarge the size of the training set to ensure the success of deep learning models in fake news detection. 

\item Adopting reinforcement learning techniques, the proposed framework {WeFEND} has the ability of selecting high-quality samples, which further leads to the improvement of the fake news detection performance.

\item We empirically show that the proposed framework {WeFEND} can effectively identify fake news and significantly outperform the state-of-the-art fake news detection models on a large-scale dataset collected from WeChat official accounts.
\end{itemize} 

\section{Related Work}
In this section, we briefly review the work related to the proposed model. We mainly focus on the following two topics: fake news detection and reinforcement learning.

There are many tasks related to fake news detection, such as rumor detection~\cite{jin2014news} and spam detection~\cite{shen2017discovering}.
Following the previous works~\cite{ruchansky2017csi,shu2017fake}, we specify the definition of fake news as news which is intentionally fabricated and can be verified as false.
Many fake news detection algorithms try to distinguish news according to their features, which can be extracted from social context and news content. 

\emph{Social context features} represent the user engagements of news on social media~\cite{shu2017fake} such as the number of followers, hash-tag(\#) retweets and the network structure\cite{wu2015false}. However, social context features can only be extracted after an accumulated period of time, and thus cannot be used in a timely detection of newly emerged fake news. 

\emph{News content features} are statistical or semantic features extracted from the textual content of news, which has been explored in many literatures of fake news detection~\cite{gupta2014tweetcred,kwon2013prominent,castillo2011information}. It is difficult to design hand-crafted textual features for traditional machine learning based fake news detection models.
To overcome this limitation, Ma et al.~\cite{ma2016detecting} and Wang et al.~\cite{wang2018eann} proposed  deep learning models to identify fake news based on news text and  multi-modal data respectively. These models have shown the improvement in detection performance, but the power of deep learning models are not yet fully unleashed due to the lack of fresh high-quality samples for training.

The manual labeling of news articles is expensive to obtain, so user feedback, which is valuable signal for fake news detection, should be incorporated into the detection process. However, feedback by users may not be reliable and user reliability is unknown. The power-law distribution of user participation rates makes it difficult to correctly estimate users' reliability~\cite{moore2008evaluating,chia2011re}. In recent work~\cite{tschiatschek2018fake,kim2018leveraging}, such crowd signals were used in fake news detection. From those news articles that are flagged by users as potential fake news, these methods select a small subset and send them to experts for confirmation. Therefore, they still require manual labeling, and valuable feedback comments are not taken into consideration.

In fact, when users flag the suspicious articles, the social media platforms usually require users to provide brief explanations on why the news is fake. Such comments provide important information for the fake news detection task. Different from the existing works,  we propose to incorporate these informative explanations into the detection model. More specifically, the proposed framework {WeFEND} can leverage report messages as weak supervision to guide fake news detection based on news content.

To further improve the detection performance, we incorporate reinforcement learning techniques~\cite{sutton1998introduction} into the proposed framework to select high-quality samples. Reinforce learning was adopted in \cite{feng2018reinforcement,wu2018reinforced} to learn an instance selector based on the prediction probabilities, but our selection approach is quite different. As the distribution of news articles is changing over time, prediction probabilities are not suitable as evaluation criteria for data selection in our problem setting. Therefore, to select high quality data for fake news detection, we propose a novel performance-driven data selection method based on reinforcement learning mechanism. 

\begin{figure}[!htb]
\centering
\includegraphics[width=2.3in]{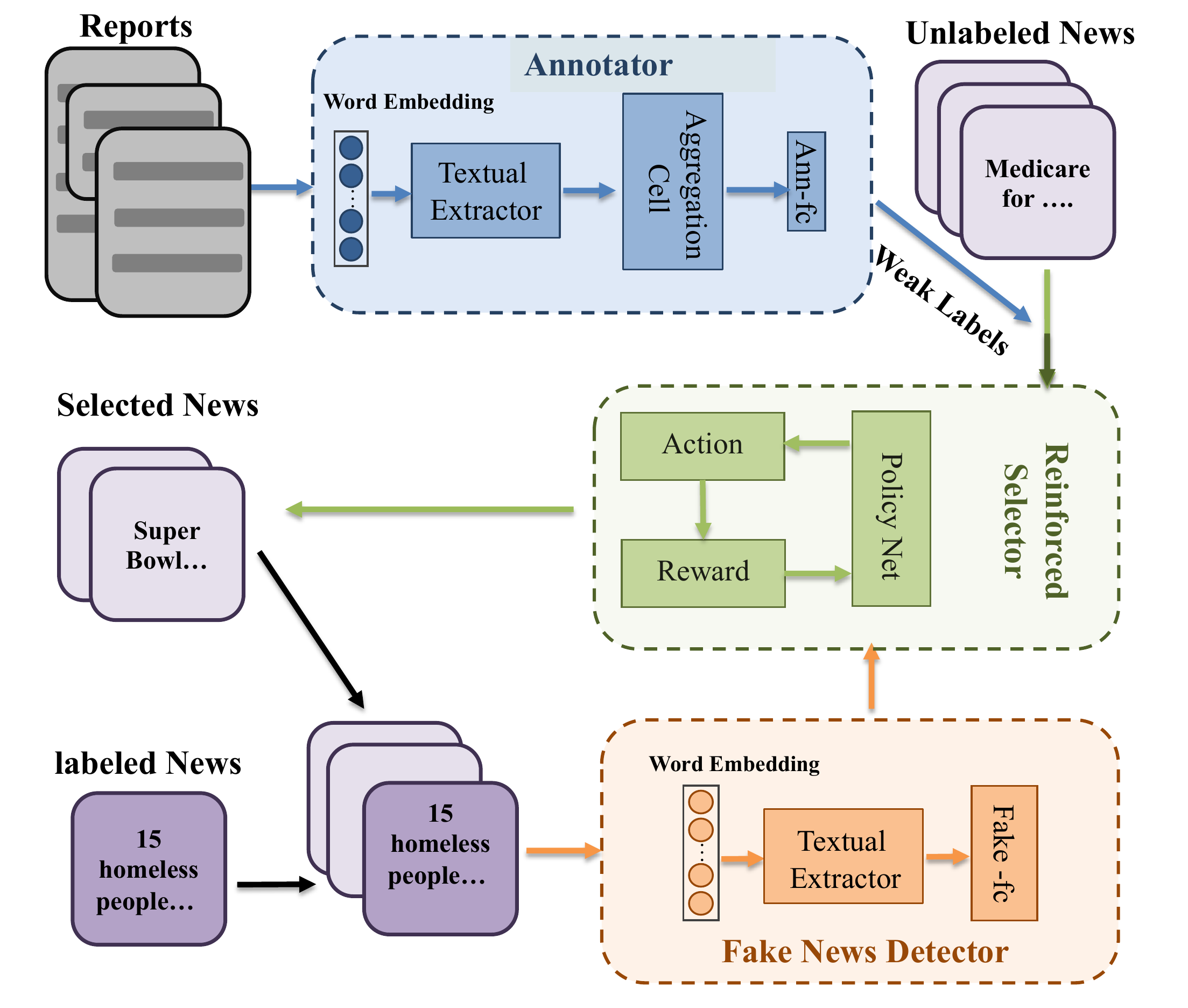}
\centering
 \caption{The architecture of proposed framework {WeFEND} which consists of annotator , reinforced selector and fake news detector.}\label{Fig:framwork}
\end{figure}

\section{Methodology}
In this section, we first briefly introduce the overview of the proposed fake news detection framework {WeFEND}, and then demonstrate each component in detail.

\subsection{Overview}
The problem setting is as follows. Each sample consists of both news articles and user feedback comments. Both are texts, and are transformed into vector representations by word embedding.  User feedback comments are referred to as \emph{reports}, which are detailed reasons and evidence provided by users about the credibility of the corresponding news articles. A small set of samples are labeled by experts as fake or real, and our objective is to predict the labels of the unlabeled samples.

Figure~\ref{Fig:framwork} shows the overview of the proposed  framework {WeFEND}. There are three key components: annotator, data selector and fake news detector. Annotator can be seen as a pretrained model on the reports with their labels. Based on the pretrained model, we can assign weak labels for the unlabeled samples according to the annotator on the reports. However, it is hard to guarantee the quality of weak labels. To automatically choose high-quality samples, we design a data selector by exploiting reinforcement learning techniques on the samples labeled by the annotator. Finally, the selected samples and the original labeled samples are used to train fake news detector. In both annotator and fake news detector, a textual feature extractor is used to extract features from input text. The details of these components are introduced in the following subsections.

\subsection{Textual Feature Extractor}
\label{section:ft}
From Figure~\ref{Fig:framwork}, we can observe that textual feature extractor is a basic module of annotator and fake news detection, as not all words are relevant to the task of fake news detector. In this paper, we choose convolutions neural network~\cite{kim2014convolutional}, which is proven effective in the fake news detection~\cite{wang2018eann}, as textual feature extractor. The input of the textual feature extractor is news content or a report message, and both can be represented as a sequential list of words. For the $t$-th word in the sentence, we represent it by the corresponding $d$ dimensional word embedding vector, denoted as $\mathbf{x}_t \in \mathbb{R}^d$, which is the input to the convolutions neural network. Details of CNN module~\cite{kim2014convolutional} are in the Supplemental Material.

The learned representation from textual feature extractor are the input features to annotator and fake news detector. Next, we will introduce how  to train an annotator and use it to assign weak labels to the unlabeled samples.

\subsection{Automatic Annotation based on Reports}
One benefit of the proposed framework is that it can automatically assign weak labels to the unlabeled news samples, which helps enlarge the size of the training set with little costs. To train such a model, we propose to use report messages provided by users as weak supervision. 

\textbf{Aggregation Cell.} One news article may have reports from multiple users, so we propose to aggregate features obtained from different reports for one sample. Since the report messages from multiple users for one piece of news are permutation invariant, we design an aggregation cell consisting of a commutative aggregation function and a fully-connected layer. The commutative aggregation function, like sum, mean and max-pooling, can combine the permutation invariant input set. We take the $i$-th sample as an example, and the $j$-th report message is represented as $r^{(i)}_j$. The corresponding report message set is denoted as $R^{(i)} = \{r^{(i)}_1, r^{(i)}_2, ..., r^{(i)}_{|R^{(i)}|}\}$, where $|R^{(i)}|$ is the number of report messages of the $i$-th sample. The report message $r^{(i)}_j \in R^{(i)}$ is first fed into the textual feature extractor to obtain an informative textual feature representation, denoted as $\mathbf{h}^{(i)}_{j}$. Then we use the aggregation cell to combine the report message set $R^{(i)}$ to learn the hidden feature representation $\mathbf{h}^{(i)}$. In order to stabilize the training procedure, we use average operation as the commutative aggregation function. The procedure of aggregation cell is represented as:
\begin{equation}
\mathbf{h}^{(i)} =  \mathtt{ReLU} (\mathbf{w}_{r} \cdot \sum_{j = 1}^{|R^{(i)}|}\frac{ \mathbf{h}^{(i)}_{j}}{|R^{(i)}|}),
\end{equation}
where $\mathbf{w}_{r}$ is the weight of the fully-connected layer. 

We feed $\mathbf{h}^{(i)}$ into the fully connected layer, denoted as Ann-fc, to output the corresponding probability of the $i$-th sample being a fake one, which is denoted as $D_r(R^{(i)}, \theta_r)$,  where $\theta_r$ represents all the parameters of the annotator and corresponding textual feature extractor. The entire report message dataset is represented as $R = \{R^{(1)}, R^{(2)}, ..., R^{(|R|)}\}$, and the corresponding ground truth labels of news are denoted as $Y = \{y^{(1)}, y^{(2)}, ..., y^{(|R|)}\}$, where $|R|$ is the number of report sets. Based on $R$ and $Y$, the loss function for the proposed annotator is defined by cross entropy as follows:

\begin{equation}
    \begin{aligned}
     L_r(R, Y; \theta_r) = &-\frac{1}{|R|} \sum_{i=1}^{|R|} [y^{(i)}\log D_r(R^{(i)}; \theta_r) \\
 &+ (1-y^{(i)})\log(1- D_r(R^{(i)}; \theta_r))].
 \end{aligned}
\end{equation}

Given the unlabeled news set $X^{u}$ with corresponding report messages, we use the trained annotator to predict their labels, which are denoted as $\hat{Y}^u$. By the annotator, we can obtain a large weakly labeled dataset $\{X^u, \hat{Y}^u\}$. However, the labels in this automatically-annotated dataset are unavoidably noisy and directly adding these samples to the training set may degrade the detection performance. 
Thus, the challenge here is how to select high-quality samples from this set to guarantee the detection performance.
To address this challenge, we propose to employ reinforcement learning techniques in the design of a data selector. The details of the proposed data selector are introduced in the following subsection.

\subsection{Data Selection via Reinforcement Learning}
\label{section:rl}

The objective of the data selector is to automatically select high-quality samples from those with weak labels obtained from the annotator. The criteria of the selection is based on whether adding the chosen sample can improve the fake news detection performance. 
According to this criteria, we design a performance-driven data selection method (called reinforced data selector) using reinforcement learning mechanism. Next, we first introduce the input data of the designed data selector, and then present the details of this data selector.

Let $\tilde{X}$ denote all the input data of the proposed reinforced data selector. However, instead of directly putting the entire dataset $\tilde{X}$ into the selector, we divide the whole dataset into $K$ small bags of data samples, i.e., $\tilde{X} = \{\tilde{X}^{(k)}\}_{k=1}^K$. A bag of data samples is  the input of the selector each time. For the $k$-th bag of data $\tilde{X}^{(k)}$, it contains $B$ samples, i.e., $\tilde{X}^{(k)} = \{x^{(k)}_1, x^{(k)}_2,..., x^{(k)}_{B}\}$. The benefit of using multiple small bags of samples is that this approach can provide more feedback to the selector and this makes the training procedure of reinforcement learning more efficient.

\textbf{Problem Formulation.} In the data selection procedure, the samples in one bag are sequentially fed into the designed reinforced data selector. For every sample, the \textit{action} of reinforced data selector is to retain or remove. The decision of the current sample $x^{(k)}_i$ is based on its  \textit{state} vector and all the previous decisions of samples $\{x^{(k)}_1, x^{(k)}_2,..., x^{(k)}_{i-1}\}$. Thus, the data selection problem can be naturally cast as a Markov Decision Process (MDP) that is the prerequisite of reinforcement learning. Since the goal of data selection is to improve the performance of fake news detection, we directly use the performance changes of fake news detection as the \textit{reward} for reinforced selector. The performance is evaluated by accuracy. For this sequential decision procedure, the reward is delayed because it can only be obtained after all the decisions are made. To solve the delayed reward problem, we employ the policy-based reinforcement learning mechanism. Since the reinforced selector needs to use the performance changes of fake news detection as reward, we introduce the fake news detector first.

\textbf{Fake News Detector.} Fake news detection model is a neural network, which consists of a textual feature extractor and a fully-connected layer, namely Fake-fc, with corresponding activation functions. The input to fake news detector is news content, and the output is the probability of the given news being fake. The detector is denoted as $D_n(\cdot; \theta_n)$, where $\theta_n$ represents all the parameters.

After introducing fake news detector, we will introduce the concepts of \textit{state}, \textit{action}, and \textit{reward} used in the proposed reinforced selector in detail as follows.

\textbf{State.} The state vector of the sample $x_i^{(k)}$ is denoted as $s_i^{(k)}$. Since every action is made based on the current sample and the chosen sample, the state vector mainly consists of two components: the representation of the current sample and the average representation of the chosen samples. 
The representation of the current sample is related to data quality and diversity. We use the output probability from the proposed annotator and the output probability of fake news detector to measure the quality of the data. To represent the data diversity, we first calculate cosine similarity between the current sample and all the chosen samples. Here each sample is represented by a vector obtained from textural feature extractor. We then select the max value of cosine similarity as the diversity. To balance the distribution of classes, the weak label of the current sample is also used as a part of the representation. Therefore, the current state vector contains four elements: 1)~the output probability from the annotator, 2)~the output probability from fake news detector, 3)~the maximum of cosine similarity between the current sample and the chosen samples, and 4)~the weak label of the current sample. The representations of all the chosen samples are defined as the average of all the chosen samples' state vectors. The concatenation of the current state vector and the average of previous state vectors is considered as the final state vector $s_i^{(k)}$.

\textbf{Action.} The action value of the reinforced selector for every sample is $1$ or $0$. $1$ represents the action to \textit{retain} the sample, and $0$ denotes the action to \textit{remove} it. To determine the action, we train a policy network, denoted as $P(\cdot; \theta_s)$, where $\theta_s$ represents the parameters. The policy network includes two fully connected layers with corresponding activation functions. Take the sample $x_i^{(k)}$ as an example. The policy network outputs a probability of retaining, denoted as $p_i^{(k)}$, based on the sample's state vector $s_i^{(k)}$:

\begin{equation}
P(s_i^{(k)}; \theta_s) = \delta(\textbf{w}_{s2} \cdot \mathtt{ReLU}(\textbf{w}_{s1}\cdot s_i^{(k)})),
\end{equation}
where $\mathbf{w}_{s1}$ and $\mathbf{w}_{s2}$ are the weights of two fully-connected layers and $\delta$ represents the Sigmoid activation function. 
Then the action $a_i^{(k)}$ is sampled according to the output probability. The policy  $\pi_{\theta_s}(s_i^{(k)}, a_i^{(k)})$ can be represented as

\[ \pi_{\theta_s}(s_i^{(k)}, a_i^{(k)})  = 
  \begin{cases}
    p_i^{(k)}       & \text{if } a_i^{(k)} = 1\\
    1 - p_i^{(k)}  & \text{if } a_i^{(k)} = 0
 \end{cases}.
\]
\normalsize
\textbf{Reward.} Since the goal of the action is to retain the samples that can bring improvement to fake news detection, we use the performance changes of detection model $D_n(\cdot; \theta_n)$ as the reward function. Given the $k$-th bag of data $\{x_1^{(k)}, x_2^{(k)},..., x_B^{(k)}\}$, the actions of retaining or removing are made based on the probability output from the policy network. 
To evaluate the performance changes, we need to set a baseline accuracy $acc$. Towards this end, we first extract a validation dataset from the whole labeled dataset. Note that all the trained model will test on this extracted validation dataset. We then calculate the baseline accuracy $acc$ with the detection model $D_n(\cdot; \theta_n)$. Since the designed data selector can choose some high-quality  samples from $\{x_1^{(k)}, x_2^{(k)},..., x_B^{(k)}\}$, the fake news detection model will be retrained using the retained data samples. A new accuracy $acc_k$ can be obtained with the retrained model on the validation dataset. Finally, the reward $R_k$ for $k$-th bag data  $\{x_i^{(k)}\}_{i=1}^B$ is represented by the difference of $acc_k$ and $acc$ as follows:
\begin{equation}
R_k = acc_k - acc.
\end{equation}

For the $k$-th bag of data $\{x_i^{(k)}\}_{i=1}^B$, we aim to maximize the expected total reward. Since the scale of $R_k$ is small, we use the summation of reward to define the objective function in order to make the training procedure more efficient. The objective function is defined as 
\begin{equation}
J(\theta_s) =  \sum_{i=1}^B\pi_{\theta_s}( s_i^{(k)}, a_i^{(k)})R_k .\\ 
\end{equation}

The derivative of the objective function above is

\begin{equation}
\begin{aligned}
\bigtriangledown_\theta J(\theta_s) &=   \sum_{i=1}^B R_k \bigtriangledown_{\theta_s} \pi_{\theta_s}(s_i^{(k)}, a_i^{(k)})\\
&=\mathbb{E}_{\pi_{\theta_s}}[ \sum_{i=1}^B R_k \bigtriangledown_{\theta_s} \log\pi_{\theta_s}(s_i^{(k)}, a_i^{(k)})]
\end{aligned}
\end{equation}

According to the policy-based reinforcement learning algorithm~\cite{sutton2000policy,sutton1998introduction}, we update the parameters $\theta$ of the policy network by stochastic gradient ascent as follows: 
\begin{equation}
\label{eq:policy_update}
{\theta_s}  \gets  {\theta_s} +\alpha   \sum_{i=1}^B R_k \bigtriangledown_{\theta_s} \log\pi_{\theta_s}(s_i^{(k)}, a_i^{(k)}),
\end{equation}
where $\alpha$ is the learning rate. To improve the exploration and stabilize training, we  train a target policy network $P(\cdot, {{\theta_s}'})$ that updates much slower than the policy network $P(\cdot, {\theta_s})$:
\begin{equation}
\label{eq:target_update}
 \theta'_s = (1-\tau) \theta'_s + \tau \theta_s.
\end{equation}

In the training stage, half of the bags are fed into the policy network $P(\cdot; \theta_s)$ and the another half of bags are fed into the target policy network $P(\cdot; \theta'_s)$. 
The detailed steps of the data selection component are summarized in Algorithm~\ref{alg:rl}. The procedure of the entire framework is introduced in the next subsection.

\vspace{-0.05in}
\begin{algorithm}
  \caption{\bf The algorithm of the reinforced selector.}\label{alg:rl}
  \flushleft
  \begin{small}
  {\bf Input:} The automatically-annotated news set $\{X^u, \hat{Y}^u\}$, the bag number $K$, the bag size $B$ and the learning rate $\alpha$\\
	\begin{algorithmic}[1]
     \State Sample $K$ bags of data $\{\tilde{X}^{(k)}, \tilde{Y}^{(k)}\}_{k=1}^K$ from $\{X^u, \hat{Y}^u\}$ and the size of every bag is $B$
    \For{ k $\in$ K}
        \If{k is even}
           \State Get the actions from policy network $P(\cdot, {\theta_s})$
        \Else
            \State Get the actions from target policy network $P(\cdot, {\theta'_s})$
        \EndIf
        \State Update the policy network   $\theta_s$ according to Eq. ~\ref{eq:policy_update}
         \State Update the target network  $\theta'_s$ according to Eq. ~\ref{eq:target_update}
    \EndFor
    \\
     \State Sample $K$ bags of data $\{\tilde{X}^{(k)}, \tilde{Y}^{(k)}\}_{k=1}^K$ from $\{X^u, \hat{Y}^u\}$ and the size of every bag is $B$
    \For{ k $\in$ $K$}
       \State  Based on actions of policy network $P(\cdot; \theta_s)$, the selected samples from data bag $\{\tilde{X}^{(k)}, \tilde{Y}^{(k)}\}$ to form a new data bag $\{X_s^{(k)}, Y_s^{(k)}\}$ 
    \EndFor
	\end{algorithmic}
	{\bf Output:} The selected data set $\{X_s, Y_s\} = \{X_s^{(k)}, Y_s^{(k)}\}_{k=1}^K$ 
  \end{small}
\end{algorithm}

\subsection{Reinforced Weakly-supervised Fake News Detection Framework}
In this subsection, we introduce how to integrate the three key components: annotator, fake news detector and reinforced selector. First, we pretrain the annotator using the labeled report data $\{R, Y\}$ and assign weak labels $\hat{Y}^u$ to the unlabeled news set $X^u$. 
The proposed reinforced selector will select high-quality  samples from the weakly labeled dataset $\{X^u, \hat{Y}^u\}$. Here we set the selected bags as $K$.  Then both the selected data set $\{X_s, Y_s\} = \{X_s^{(k)}, Y^{(k)}_s\}_{k=1}^K$ and the original labeled data are fed into the fake news detector for training. Thus, the final loss of fake news detection  consists of two sub losses:
\begin{equation}
\begin{aligned}
\label{eq:fake_news_loss2}
 L_n(X, Y,X_s, Y_s; \theta_n) =  &\lambda_l \cdot L_n^l(X, Y; \theta_n) \\&+ \lambda_s \cdot L_n^s(X_s, Y_s; \theta_n),
\end{aligned}
\end{equation}
where $L_n^l(X, Y; \theta_n)$ and $L_n^s(X_s, Y_s; \theta_n)$ are the losses on a small amount of manually labeled data and automatically-annotated data set respectively. Here the $\lambda_l$ and $\lambda_u$ control the balance between $L_n^l(\theta_n)$ and $L_n^s(\theta_n)$, and we simply set the values of $\lambda_l$ and $\lambda_u$ as 1. The two losses are defined by cross entropy respectively as follows:

\begin{equation}
\begin{aligned}
L_n^l(X, Y; \theta_n) = &-\mathbb{E}_{(x, y) \sim (X, Y)}~[y\log(D_n(x; \theta_n))  \\&+ (1-y)\log(1- D_n(x; \theta_n))],
\end{aligned}
\end{equation}
\begin{equation}
\begin{aligned}
 L_n^s(X_s, Y_s; \theta_n) = &-\mathbb{E}_{(x_s, y_s) \sim (X_s, Y_s)}~[y_s\log(D_n(x_s, \theta_n)) \\ &+(1-y_s)\log(1- D_n(x_s; \theta_n))].
\end{aligned}
\end{equation}
The detailed steps of the proposed framework are summarized in Algorithm~\ref{alg:framework}.

\vspace{-0.1in}
\begin{algorithm}
  \caption{\bf Reinforced weakly-supervised fake news detection framework.}\label{alg:framework}
  \flushleft
  \begin{small}
  {\bf Input:} The labeled input with news content $\{X, Y\}$ and the corresponding report messages $\{R, Y\}$, the unlabeled input with news content $X^u$ and the corresponding report messages $R^u$ and the learning rate $\alpha$\\
	\begin{algorithmic}[1]
     \For{ number of training epochs}
        \State Update the annotator's parameters $\theta_r$: 
         \State  \hspace{\algorithmicindent} 
         $\theta_r \gets \theta_r  - \alpha \bigtriangledown_{\theta_r} L_r(R, Y; \theta_r).$
    \EndFor
    
    \State Use the trained annotator to assign weak labels $\hat{Y}^u$ to unlabeled news $X^u$ based on report messages $R^u$. 
   
     \For{number of training epochs}
        \State Select data set $\{X_s, Y_s\}$ from $\{X^u, \hat{Y}^u\}$  according to Algorithm~\ref{alg:rl};
        \State Update the fake news detector's parameters $\theta_n$:
        \State  \hspace{\algorithmicindent} 
         $\theta_n \gets \theta_n  - \alpha \bigtriangledown_{\theta_n} L_n(X, Y, X_s, Y_s; \theta_n).$
    \EndFor
	\end{algorithmic}
  \end{small}
\end{algorithm}
\section{Experiments}

In this section, we introduce the dataset used in the experiments, present the compared fake news detection models, validate the effectiveness and  explore some insights of the proposed framework.
\subsection{Dataset}

To fairly evaluate the performance of the proposed framework, we collect a dataset from WeChat's Official Accounts and conduct comprehensive experiments to analyze the performance.  This dataset includes user reports and will be publicly released in future to encourage research on fake news detection.
\begin{table}[htb]
\centering
  \caption{The Statistics of the WeChat Datasets.}
  \label{tab:stat}
  \resizebox{0.9\linewidth}{!}{\begin{tabular}{c|c|c|c|c}
    \hline
    &&\# News & \# Report  & \# Avg. Reports/News \\
     \hline
     Unlabeled&-&22981&31170&1.36 \\  
      \hline
      \multirow{2}{*}{Labeled Training}& Fake &1220 &2010 & 1.65\\ 
      \cline{2-5}
      & Real& 1220 & 1740& 1.43\\
     \hline
     \multirow{2}{*}{Labeled Testing} & Fake& 870& 1640 & 1.89 \\   
     \cline{2-5}
     & Real&870 & 1411 & 1.62\\
      \hline
     
\end{tabular}}
\end{table}
In this dataset, the news are collected from WeChat's Official Accounts, dated from March 2018 to October, 2018. To facilitate the detection fake news, the WeChat's Official Account encourages users to report  suspicious articles, and write feedback to  explain why they think these articles are suspicious. To obtain a small set of labeled samples, we first collect the news with reports and then send them to the experts of WeChat team for verification. Thus, the manually labeled fake and real news both have report messages. 
We split the fake news and real news into training and testing sets according to the post timestamp. The news in the training data were posted from March 2018 to September 2018, and testing dataset is from September 2018 to October 2018. There is no overlapped timestamp of news between these two sets. This design is to evaluate the performance of fake news detection on the fresh news. We also have an unlabeled set containing a large amount of collected news without annotation. The time window of the unlabeled set is from September to October 2018. The detailed statistics are shown in the Table~\ref{tab:stat}. 
Note that the headlines can be seen as the summary of the news content. In the manual annotation process, experts only look at headlines to conduct labeling. Thus, in this paper, we  use headlines as the input data. \\

\noindent\textbf{Settings.}

$\bullet$ \textbf{Supervised Setting.} We split the manually labeled training set into two sets in a ratio 8:2. 20\% of training set is used as a validation set to select parameters, and the remaining 80\% of data is used for training purpose. 


$\bullet$ \textbf{Semi-supervised Setting.} We still split the data as the supervised setting, but also use the unlabeled data in the semi-supervised setting. The combination of 80\% of training set and unlabeled set are used for training. We use the entropy minimization to define the loss on unlabeled data as \cite{grandvalet2005semi}. Considering the relative data size, we set the ratio between two losses on labeled set and unlabeled set as 1:0.1. 

$\bullet$ \textbf{Weakly-supervised Setting.} We pretrain the proposed annotator on \emph{reports} in the training set and use the pretrained annotator to automatically annotate the unlabeled set. Based on the annotation, we divide the unlabeled data into two sets: weakly fake data and weakly real data. For each set, we randomly select a subset of data samples. The number of data samples is 10\% of the whole unlabeled data. These two subsets consist of a new validation set, which is used for choosing the best parameters. All the remaining data with weak labels are used for training the model.


$\bullet$ \textbf{Automatically annotated.}  All the settings for Hybird are the same as those of Weakly-supervised, but in the last step, we use both the labeled training data (80\% as the Supervised and Semi-supervised settings) and the remaining data with weak labels to train the model. \\

\noindent\textbf{Baseline Approaches.}

$\bullet$ \textbf{LIWC}~\cite{pennebaker2015development}. 
LIWC stands for Linguistic Inquiry and Word
Count, which is widely used to count words in psychologically meaningful categories. 
Previous work on verbal deception
detection showed that LIWC is a valuable tool for the deception detection in various contexts~\cite{ott2011finding,perez2017automatic}. Based on LIWC features, we detect fake news with different learning algorithms including Logistic Regression ({LIWC-LR}), SVM~({LIWC-SVM}) and Random Forest~({LIWC-RF}). The algorithms are implemented by scikit-learn machine learning framework~\cite{scikit-learn} in Python.

$\bullet$ \textbf{EANN}~\cite{wang2018eann}.
{EANN} is one of the state-of-the-art models for fake news detection.
It consists of three components: feature extractor, event discriminator and fake news detector. In this work, our input is only text. Thus, we remove the image feature extractor and keep textual extractor. The textual extractor is based on CNN that has the identical architecture with our CNN textual feature extractor.
We follow the settings mentioned in~\cite{wang2018eann} to cluster the headlines to get event id. 

$\bullet$ \textbf{CNN}.
{CNN} model employs Convolutional Neural Network to extract a textual feature vector for each headline as illustrated in the subsection~\ref{section:ft}. The extracted textual feature is fed into one fully connect layer to adjust the dimensionality. Built on top of CNN extractor, a fully connected layer with softmax function is used to predict whether this headline is fake or not. 
For textual feature extractor, we use 40 filters with window size ranging from 1 to 6, and the fully connected layer adjusts the dimensionality of features from 240 to 40.  The last fully connected layer take 40-dimensional feature vector as input to identify whether the news is fake or not. 


$\bullet$ \textbf{LSTM}.
{LSTM} uses one-layer LSTM as textual feature extractor, which is illustrated in the subsection~\ref{section:ft}. The latent representations are obtained by averaging the outputs of RNN at each timestep, and then these representations are fed into a fully connected layer to make predictions. The hidden size of LSTM is set as 40. Built on top of LSTM extractor, a fully connected layer outputs the probability of fake news. 

Furthermore, 
the complete {WeFEND} model consists of three components: annotator, fake news detector and data selector. To show the role of data selector, we design one variant of the
proposed model named  {WeFEND}$-$ , which does not include data selector.\\

\noindent \textbf{Implementation Details.} The 200 dimensional pre-trained word-embedding weights~\cite{song2018directional} are used to initialize the parameters of the embedding layer. The architecture of the detector is the same as that of the baseline CNN.
In the \emph{annotator}, the weight $\mathbf{w}_r \in \mathbb{R}^{40 \times 20}$.  In the \emph{reinforced selector},  $\mathbf{w}_{s1} \in \mathbb{R}^{8  8}$ and $\mathbf{w}_{s2} \in \mathbb{R}^{8 \times 1}$. We set the bag size $B$ same as mini-batch size, $\tau = 0.001$ and $K = 200$.
We implement all the deep learning baselines and the proposed framework with PyTorch 1.2. For training models, we use Adam~\cite{kingma2014adam} in the default setting. The learning rate $\alpha$ is 0.0001.  We use mini-batch size of 100 and training epochs of 100. 

\subsection{Performance Comparison}

\begin{table*}[htb]
\centering
\caption{The performance comparison of different methods on WeChat dataset.}
\label{tab: expermental_results}
\resizebox{0.9\linewidth}{!}{
\begin{tabular}{c|c|c|c|c|c|c|c|c|c}
\cline{1-10}
\multirow{2}{*}{Category} &
  \multirow{2}{*}{Method} & \multirow{2}{*}{Accuracy}& \multirow{2}{*}{AUC-ROC}&
\multicolumn{3}{c|}{Fake News}
&
\multicolumn{3}{c}{Real News} \\
\cline{5-10}
&&&& {Precision}  & {Recall} & {F$_1$} &{Precision}  & {Recall} & {F$_1$} \\
\cline{1-10}
\multirow{6}{*}{Supervised}&{LIWC-LR} & 0.528 & 0.558 & 0.604 & 0.160 & 0.253 & 0.517& 0.896 & 0.655  \\
&{LIWC-SVM} & 0.568 & 0.598 & 0.574 & 0.521 & 0.546 & 0.563& 0.614 & 0.587  \\
&{LIWC-RF} & 0.590 & 0.616 & 0.613 & 0.483 & 0.541 & 0.574& 0.696 & 0.629  \\
&{LSTM} & 0.733 & 0.799 & 0.876 & 0.543 & 0.670 & 0.669& \textbf{0.923} & 0.775  \\
&{CNN} & 0.747 & 0.834 & 0.869
& 0.580 & 0.696 & 0.685& 0.913 & 0.783  \\
&{EANN}& 0.767& 0.803 & 0.863 & 0.634 & 0.731 & 0.711 & 0.899 & 0.794\\
\cline{1-10}
\multirow{2}{*}{Semi-supervised}&{LSTM}$_{semi}$& 0.753 & 0.841 & 0.854 & 0.611 & 0.713 & 0.697 & 0.895 & 0.784   \\
&{CNN}$_{semi}$& 0.759 & 0.848&0.850 & 0.630 & 0.723 & 0.706 &0.889&0.787   \\
\cline{1-10}
\multirow{2}{*}{Weakly supervised}&{LSTM$_{weak}$}& 0.762 & 0.813 & 0.804 & 0.692 & 0.744 & 0.730 & 0.831 & 0.777   \\
&{CNN$_{weak}$}& 0.759 & 0.823&0.754 & 0.769 & 0.762 & 0.766 &0.749&0.757   \\
\cline{1-10}
\multirow{2}{*}{Automatically annotated}&{WeFEND}$-$ &  0.807 & 0.858 & 0.846 & \textbf{0.751} & 0.795 & 0.776 & 0.863 & 0.817  \\
&{WeFEND} & \textbf{0.824} & \textbf{0.873} & \textbf{0.880} & \textbf{0.751} & \textbf{0.810} & \textbf{0.783} & 0.898 & \textbf{0.836}  \\
\hline
\end{tabular}
}
\end{table*}

Table~\ref{tab: expermental_results} shows the performance of different approaches on the WeChat dataset. We can observe that that the proposed framework achieves the best results in terms of Accuracy, AUC-ROC, precision, recall and $F_1$ measurement.

In the supervised setting, LIWC-LR achieves the worst performance. The reason is that LIWC-LR is a linear model and hard to discriminate the complicated distributions of fake and real news content. Compared with LIWC-LR, LIWC-SVM and LIWC-RN improve the performance in terms of most measurements. However, compared with traditional machine learning models, deep learning based models, including LSTM, CNN and EANN, significantly improve the performance. This confirms that deep learning models have superior ability to extract informative features for detection. In particular, compared with the best traditional machine learning baseline LIWC-RF, CNN achieves around 27\% and 35\% improvement on Accuracy and AUC-ROC respectively. EANN model has the ability to capture the dynamic nature of news by learning the event-invariant feature representations. It leads to the performance improvement and better generalization ability compared with the plain LSTM and CNN.

Along with the setting of supervised learning, in semi-supervised setting, we incorporate external unlabeled news. We run LSTM and CNN models in the semi-supervised setting. Since the number of data largely increases, we can observe the performance improvement in both models. Take LSTM-based models as an example. The Accuracy and AUC-ROC of LSTM$_{semi}$ increases 3\% and 5\% respectively, compared with supervised LSTM. This illustrates that using unlabeled data enlarges size of training set and achieves performance improvement.

In the weakly supervised setting, we first train the proposed annotator on the reports with labels, and then use the well-trained annotator to assign  ``weak'' labels for unlabeled news according to their reports. Finally, a fake news detection model can be trained using the news content and the corresponding ``weak'' labels. In Table~\ref{tab: expermental_results},   weakly supervised CNN$_{weak}$ and LSTM$_{weak}$ have achieved better performance compared with their supervised versions. This confirms the importance of weak supervision from reports for fake news detection.

The advantage of the proposed framework is that it can automatically annotate unlabeled news. From the results shown in Table~\ref{tab: expermental_results}, we can observe that the performance of {WeFEND}$-$ is better than this of models in the supervised setting and semi-supervised setting.

Though  incorporating automatic annotation as weak supervision helps fake news detection in some aspects,  weak supervision is unavoidably noisy. In Table~\ref{tab: expermental_results}, the recall values of {WeFEND}$-$ improve as the coverage is increasing, but their precision values for fake news detection decrease. This shows that incorporating weak supervision may add more false positive examples. For real news, since the majority of unlabeled data with reports is still real news, the precision still improves. To reduce the influence of noisy labels, the proposed framework {WeFEND} has the data selector component based on reinforcement learning techniques. After incorporating data selector, the precision values of fake news and real news are improved compared with their reduced version in the same hybrid setting. Furthermore, we can observe from Table~\ref{tab: expermental_results} that the proposed {WeFEND} achieve the best performance compared with all the baselines.

\subsection{Insight Analysis}
In this subsection, we aim to answer two important questions:

$\bullet$  Does the distribution of news change with time?

$\bullet$  why should we use the reports to annotate the fake news?

\begin{figure}[hbt]
\centering
\begin{subfigure}{.23\textwidth}
\includegraphics[width=1.4in]{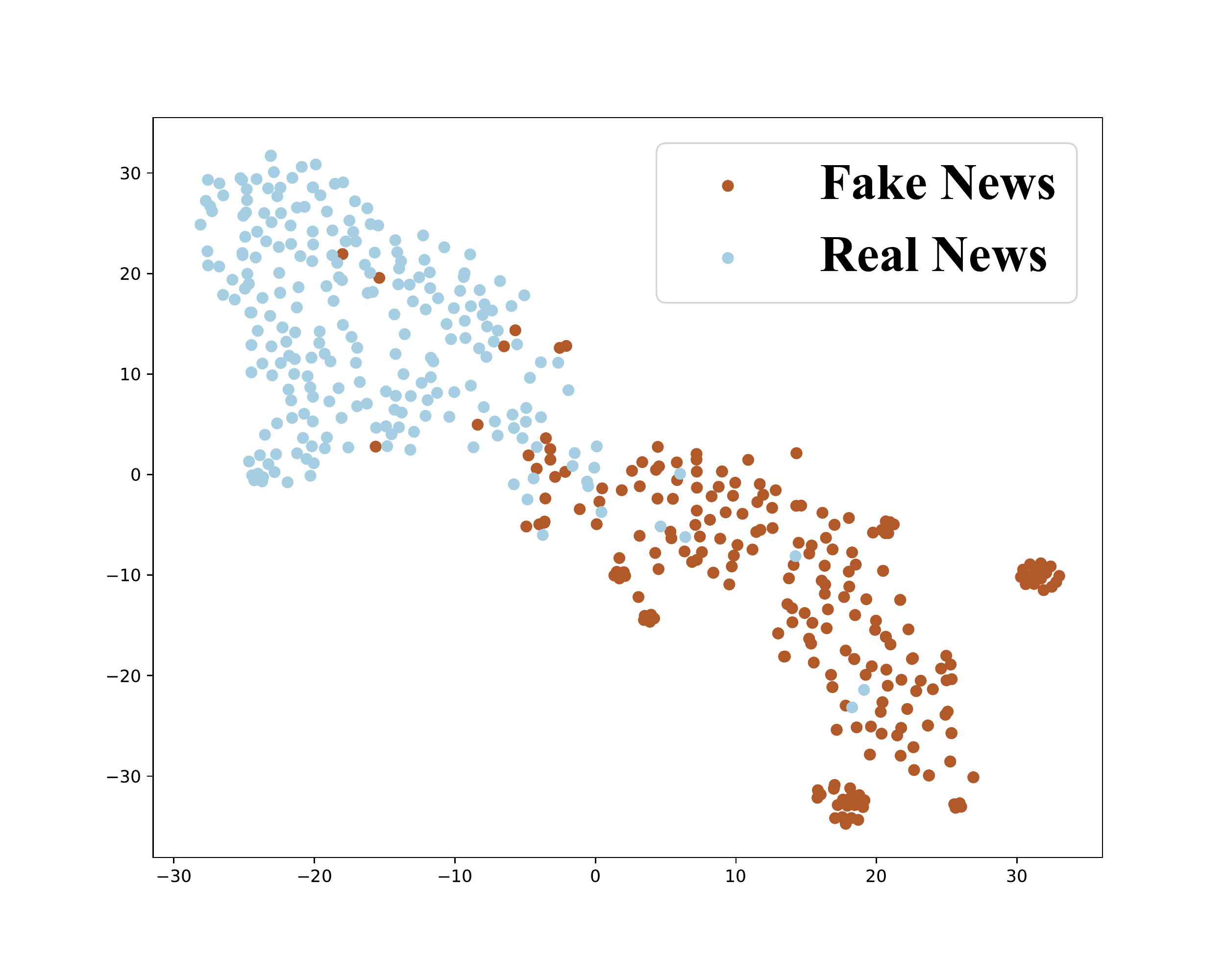}
\caption{Same Time}
\label{fig:same}
\centering
\end{subfigure}
\begin{subfigure}{.23\textwidth}
\includegraphics[width=1.4in]{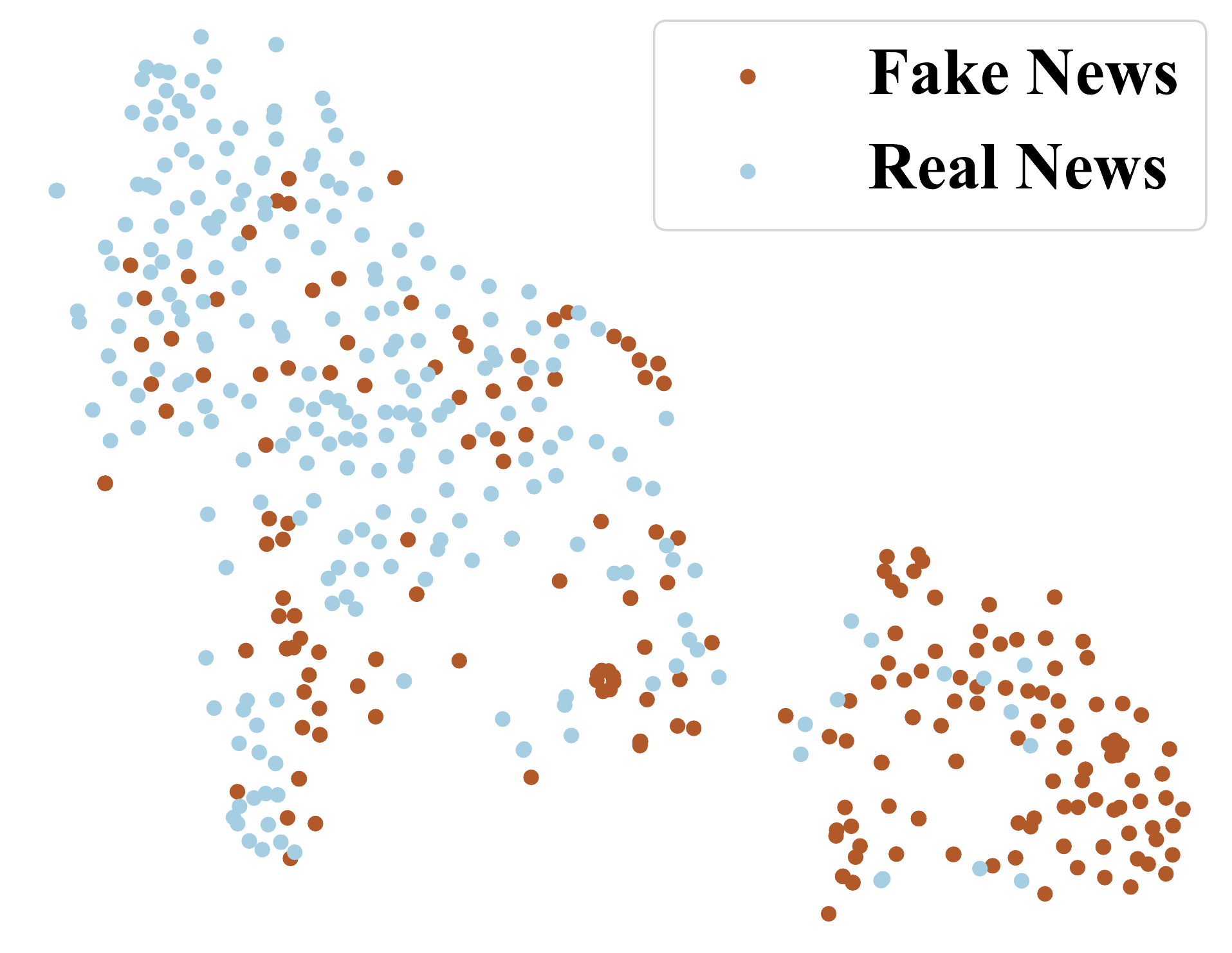}
\caption{Different Time}
\label{fig:different}
\centering
\end{subfigure}
 \caption{The Visualization of latent representations on news contents in the same time set ($D_s$) and different time set ($D_d$).}\label{Fig:different}
\end{figure}

To intuitively show the dynamic distribution of news, we conduct an experiment to compare the news in the different time windows. The experiment is designed as follows. We first split the original training dataset consists of news content and reports into two sets: 80\% data as the new training set (denoted as $D_t$) and the remaining 20\% data as the testing set for the same time window setting (denoted as $D_s$). For the different time window setting, we randomly select a subset samples from original testing dataset, which is denoted as $D_d$. The number of samples in $D_s$ is similar as that in $D_d$. The fake news detector based on CNN extractor is first trained on the news content of $D_t$, and then we separately test the model on $D_s$ and $D_d$. Figure~\ref{Fig:different} shows the visualizations of latent feature representations on $D_s$ and $D_d$ in t-SNE~\cite{maaten2008visualizing}. As can be seen, the feature representations on $D_s$ in the Figure~\ref{fig:same} are very discriminative, and the segregated area between fake and real news is clear. Compared with Figure~\ref{fig:same}, the feature representations on $D_d$ in  Figure~\ref{fig:different} are twisted together. This comparison shows the feature representations of news in different time windows are significantly different with each other.

\begin{figure}[hbt]
\centering
\begin{subfigure}{.23\textwidth}
\includegraphics[width=1.4in]{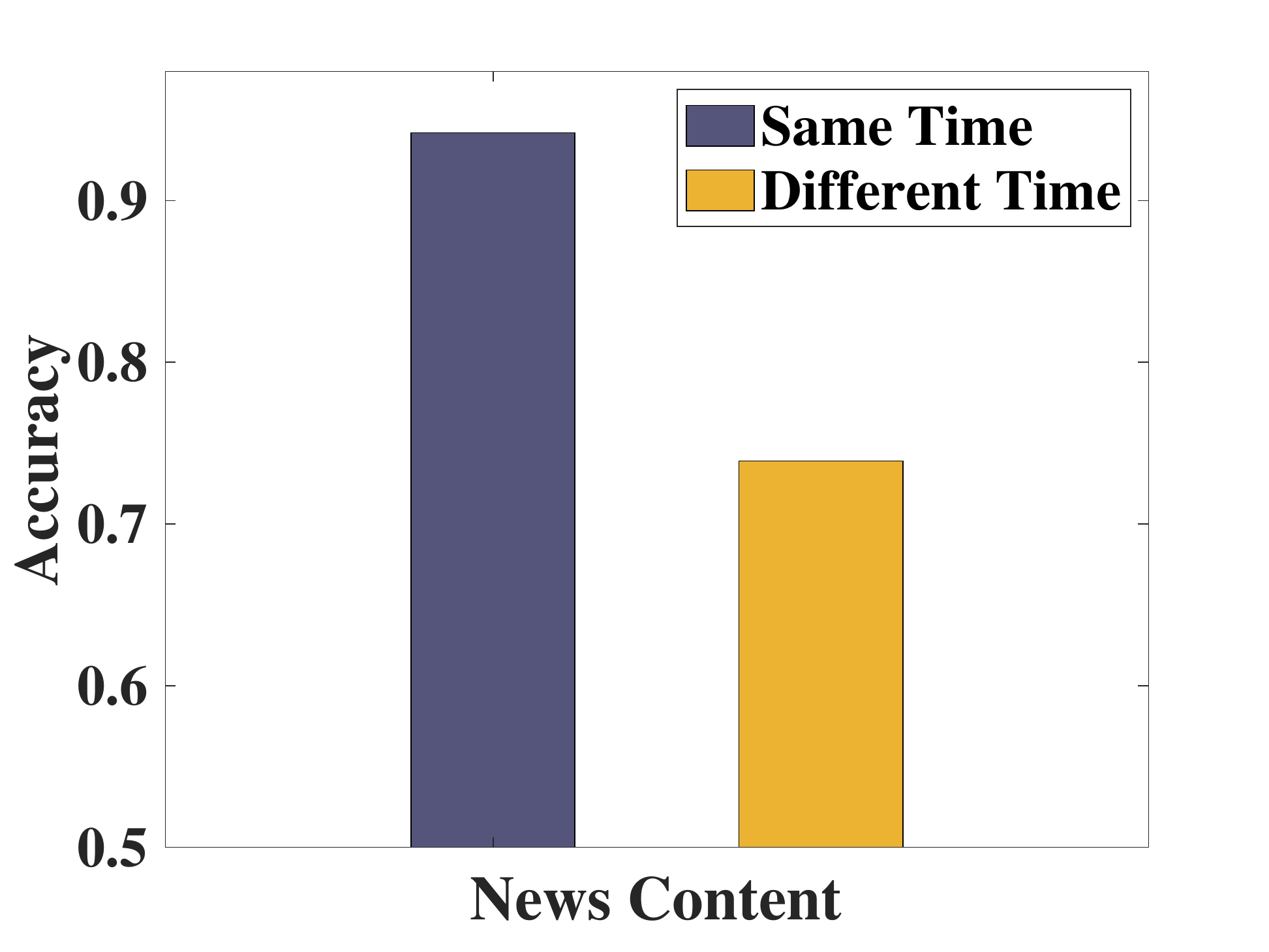}
\label{fig:subfig_train_a}
\centering
\end{subfigure}
\begin{subfigure}{.23\textwidth}
\includegraphics[width=1.4in]{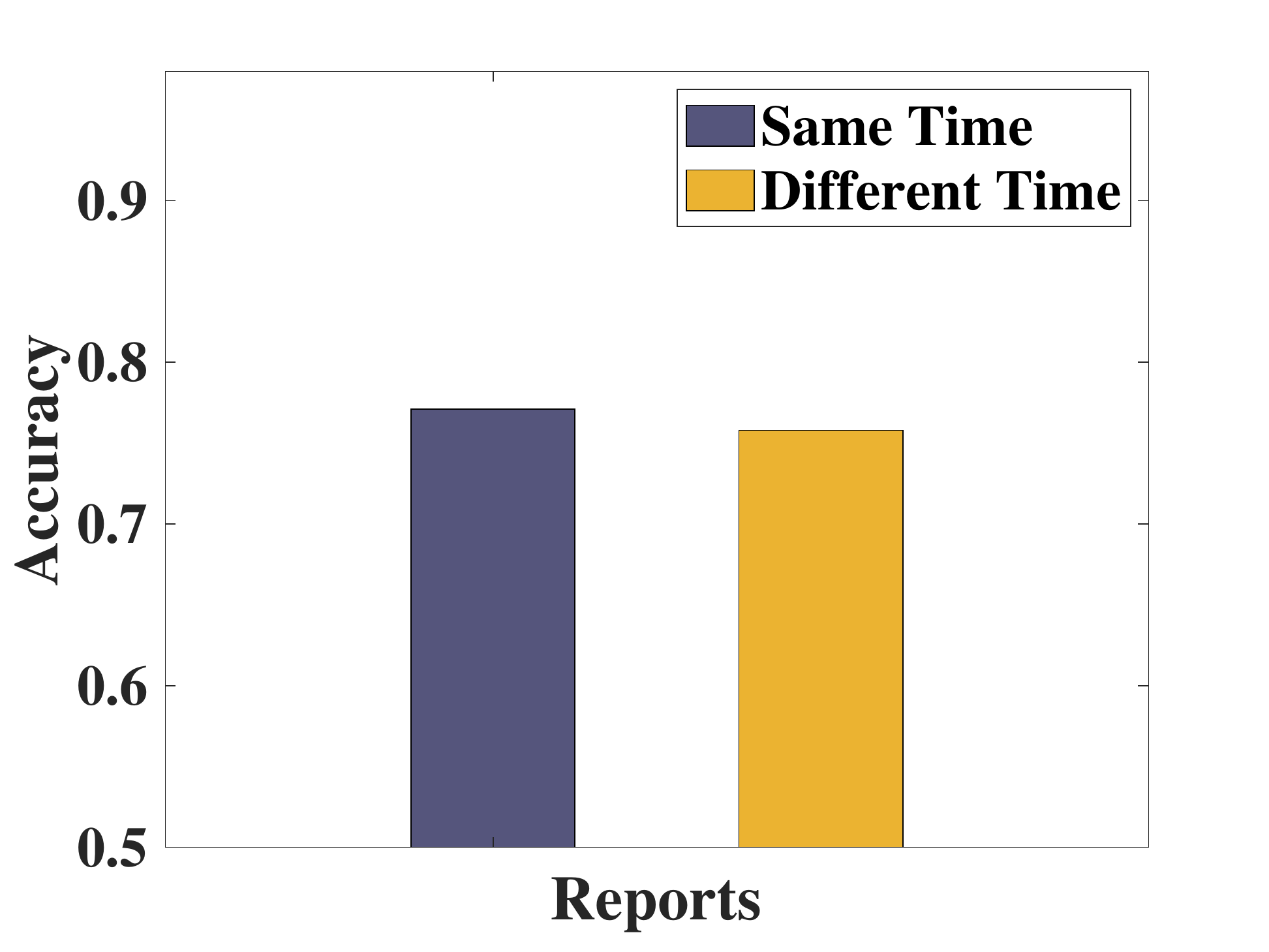}
\centering
\end{subfigure}
 \caption{The performance comparison of fake news detection on news content and reports in the two sets ($D_s$ and $D_d$).}\label{Fig:perforamcne}
\end{figure}


To further confirm the dynamic distribution, we also compare the performance of the fake news detector on the same time set $D_s$ with its performance on the different time set $D_d$. The performance comparison results are shown in the Figure~\ref{Fig:perforamcne}. As can be seen, the accuracy of detector on the same time set is around 90\%. However, for the different time set, the accuracy is only around 70\% . The significant performance difference between two sets confirms that the distribution of news is changing.

Due to the dynamic nature of news, the annotation needs to be timely to cover news articles on newly emerged events. To address this issue, we propose to use reports from users as weak supervision to automatically annotate fresh news. Next, we conduct experiments to demonstrate why  reports are useful for this purpose.  Different from the previous experiments using news content, in this experiment, we train the annotator based on reports in the set $D_t$ and test the performance on the sets $D_s$ and $D_d$. The results are shown in Figure~\ref{Fig:perforamcne}. We can observe that using report information, the annotator can achieve similar performance on the same and different time set. This demonstrates that the quality of fake news annotation on reports does not change with time.

\begin{figure}[hbt]
\centering
\begin{subfigure}{.23\textwidth}
\includegraphics[width=1.4in]{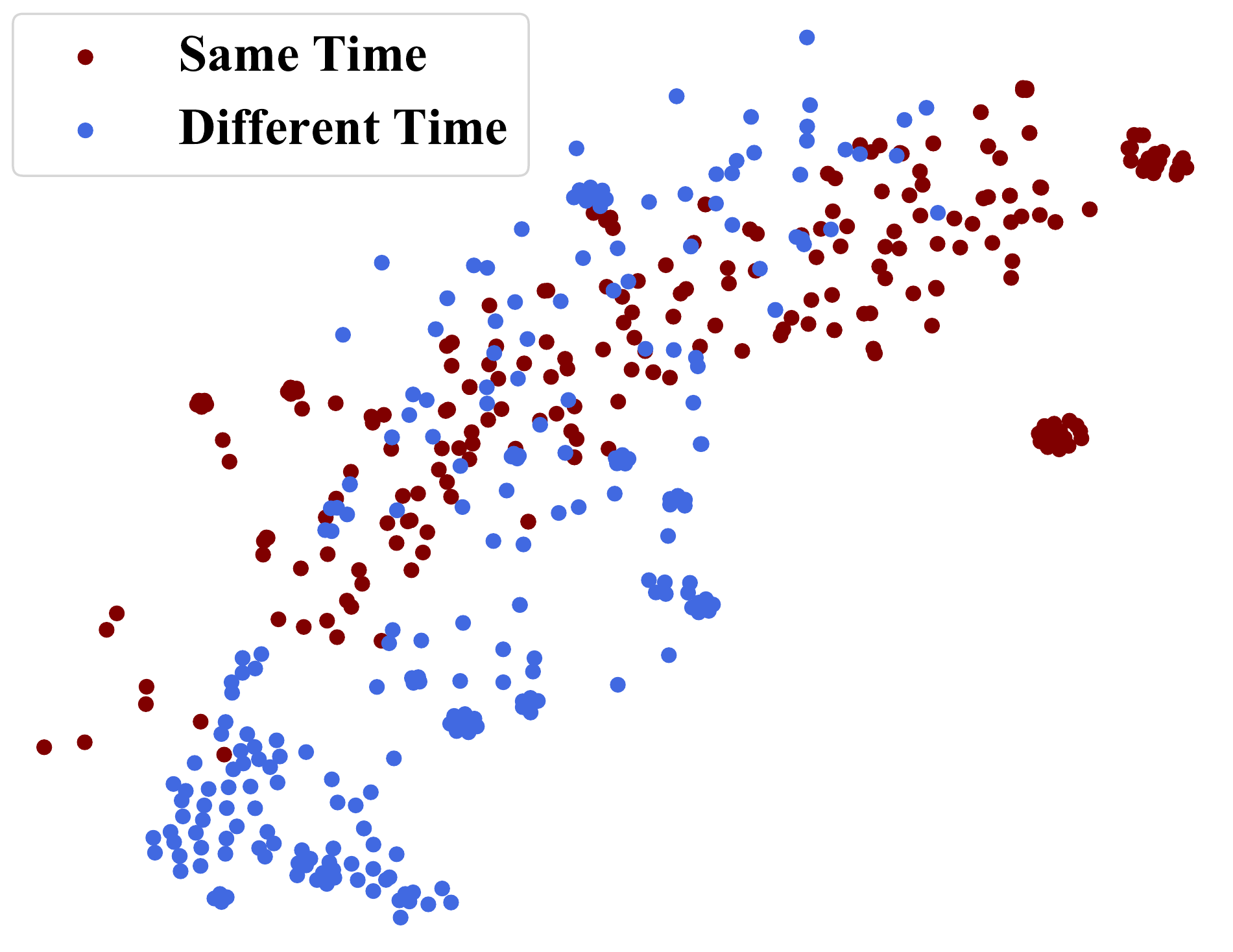}
\caption{News Content}
\label{fig:subfig_headline_fake}
\centering
\end{subfigure}
\begin{subfigure}{.23\textwidth}
\includegraphics[width=1.4in]{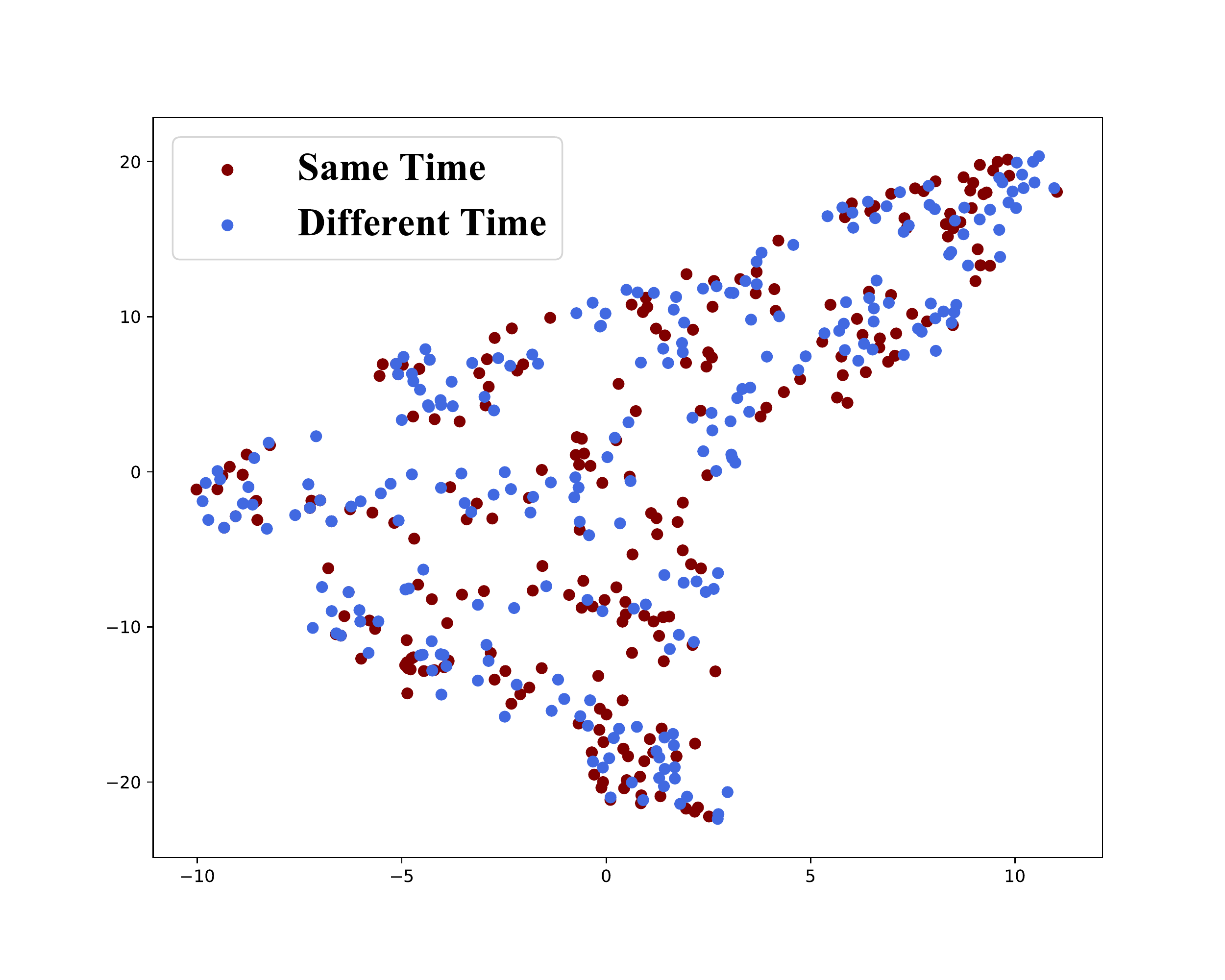}
\caption{Reports}
\label{fig:subfig_report_fake}
\centering
\end{subfigure}
 \caption{The Visualization of latent representations for news content and reports of fake news.}\label{Fig:fake_news}
\end{figure}

To further explain and confirm that annotation based on reports can achieve consistent quality, we show the distributions of reports on two time sets. Since the real news is easy to collect, the goal of annotation procedure is to expand the size of fake news samples. Thus, to analyze the distribution of reports, we mainly focus on fake news samples. The distributions of reports on the same and different time set are shown in the Figure~\ref{Fig:fake_news}. For clear comparison between the distribution of reports and news content, the feature representations of news content for fake news are also shown in Figure~\ref{fig:subfig_headline_fake}. From Figure~\ref{fig:subfig_headline_fake}, we can observe that although the distributions of news content in the same time set and different time set have  overlaps, the samples from two set are separately clustered at the top right and bottom left corner. This shows the distribution of news contents changes with time. In contrast, the feature representations of report messages from two sets are all twisted and cannot be distinguished as shown in Figure~\ref{fig:subfig_report_fake}. This proves that the distributions of reports is time invariant and further explains why the model trained on report messages achieves a consistent performance. Thus, the annotation based on reports can guarantee consistent quality even for fresh news articles.

\subsection{Importance of Reinforced Selector}
\begin{figure}[htb]
\centering
\includegraphics[width=2in]{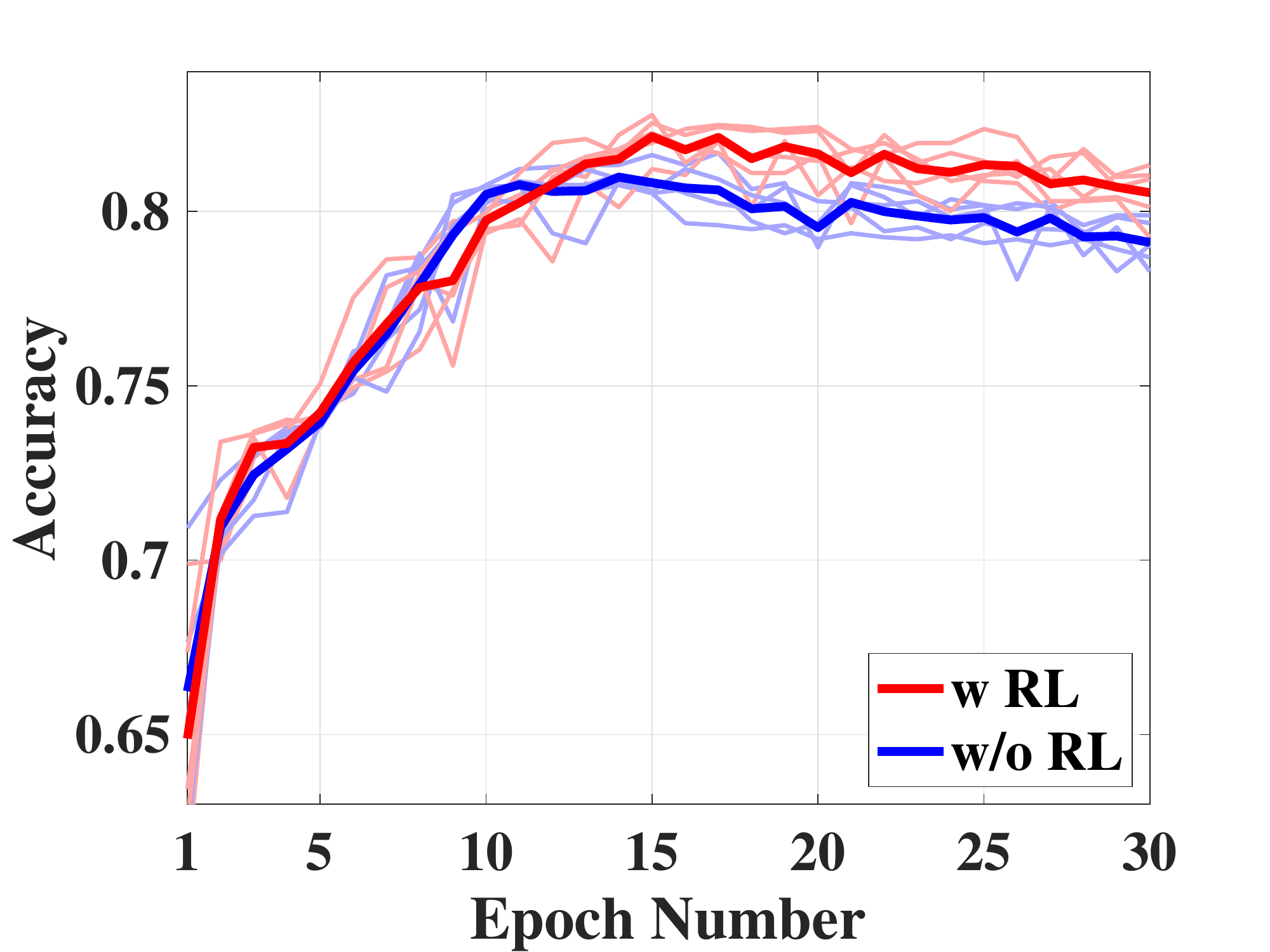}
\centering
 \caption{The changes of Accuracy in terms of the number of Epochs.}\label{Fig:rl_performance}
\end{figure}

To demonstrate the importance of reinforced selector, we run WeFEND$-$ (``w/o RL'') and WeFEND (``w RL'') 5 times, and the performance comparison during the first 30 epochs is shown in Figure~\ref{Fig:rl_performance}. Note that the only difference between two models is whether it has the component of reinforced selector or not. The solid line represents the average accuracy of 5 times, and the line with light color represents the accuracy value of a single time. As the probability output from fake news detection model can provide more information for the reinforced selector, we can observe that the average accuracy of the model with reinforced selector is stably higher than that w/o reinforced selector after 12 epochs from Figure~\ref{Fig:rl_performance}. 
The ablation study shows that the designed reinforced selector is effective in improving the performance of fake news detection.

\section{Conclusions}
 In this paper, we proposed to investigate the important problem of fake news detection. The dynamic nature of news make it infeasible to obtain continuously labeled high-quality samples for training effective models, especially for training powerful deep learning based models. Therefore, we proposed a novel framework that can leverage user reports as weak supervision for fake news detection. The proposed framework  works by integrating three components including the annotator, the reinforced selector and the fake news detector. The annotator automatically annotates unlabeled news articles as real or fake based on user reports. The reinforced selector based on reinforcement learning techniques chooses high-quality samples from those labeled by the annotator. The fake news detector then predicts the label of all the news articles by a model trained on the enhanced training set generated by both annotator and reinforced selector. By enhancing the quality and size of the training set, the proposed framework  thus has shown significantly improved performance in fake news detection. This was demonstrated in a series of experiments conducted on a WeChat dataset consisting of news articles and user feedback.

\bibliographystyle{aaai}
\bibliography{ref}

\begin{thebibliography}{}

\bibitem[\protect\citeauthoryear{Castillo, Mendoza, and
  Poblete}{2011}]{castillo2011information}
Castillo, C.; Mendoza, M.; and Poblete, B.
\newblock 2011.
\newblock Information credibility on twitter.
\newblock In {\em Proceedings of WWW},  675--684.

\bibitem[\protect\citeauthoryear{Chia and Knapskog}{2011}]{chia2011re}
Chia, P.~H., and Knapskog, S.~J.
\newblock 2011.
\newblock Re-evaluating the wisdom of crowds in assessing web security.
\newblock In {\em Proceedings of FC},  299--314.

\bibitem[\protect\citeauthoryear{Conroy, Rubin, and
  Chen}{2015}]{conroy2015automatic}
Conroy, N.~J.; Rubin, V.~L.; and Chen, Y.
\newblock 2015.
\newblock Automatic deception detection: Methods for finding fake news.
\newblock {\em Proceedings of the Association for Information Science and
  Technology} 52(1):1--4.

\bibitem[\protect\citeauthoryear{Feng \bgroup et al\mbox.\egroup
  }{2018}]{feng2018reinforcement}
Feng, J.; Huang, M.; Zhao, L.; Yang, Y.; and Zhu, X.
\newblock 2018.
\newblock Reinforcement learning for relation classification from noisy data.
\newblock In {\em Proceedings of AAAI}.

\bibitem[\protect\citeauthoryear{Grandvalet and
  Bengio}{2005}]{grandvalet2005semi}
Grandvalet, Y., and Bengio, Y.
\newblock 2005.
\newblock Semi-supervised learning by entropy minimization.
\newblock In {\em Proceedings of NIPS},  529--536.

\bibitem[\protect\citeauthoryear{Gupta \bgroup et al\mbox.\egroup
  }{2014}]{gupta2014tweetcred}
Gupta, A.; Kumaraguru, P.; Castillo, C.; and Meier, P.
\newblock 2014.
\newblock Tweetcred: Real-time credibility assessment of content on twitter.
\newblock In {\em International Conference on Social Informatics},  228--243.

\bibitem[\protect\citeauthoryear{Jin \bgroup et al\mbox.\egroup
  }{2014}]{jin2014news}
Jin, Z.; Cao, J.; Jiang, Y.-G.; and Zhang, Y.
\newblock 2014.
\newblock News credibility evaluation on microblog with a hierarchical
  propagation model.
\newblock In {\em Proceedings of ICDM},  230--239.

\bibitem[\protect\citeauthoryear{Kim \bgroup et al\mbox.\egroup
  }{2018}]{kim2018leveraging}
Kim, J.; Tabibian, B.; Oh, A.; Sch{\"o}lkopf, B.; and Gomez-Rodriguez, M.
\newblock 2018.
\newblock Leveraging the crowd to detect and reduce the spread of fake news and
  misinformation.
\newblock In {\em Proceedings of WSDM},  324--332.

\bibitem[\protect\citeauthoryear{Kim}{2014}]{kim2014convolutional}
Kim, Y.
\newblock 2014.
\newblock Convolutional neural networks for sentence classification.
\newblock {\em arXiv preprint arXiv:1408.5882}.

\bibitem[\protect\citeauthoryear{Kingma and Ba}{2014}]{kingma2014adam}
Kingma, D.~P., and Ba, J.
\newblock 2014.
\newblock Adam: A method for stochastic optimization.
\newblock {\em arXiv preprint arXiv:1412.6980}.

\bibitem[\protect\citeauthoryear{Kwon \bgroup et al\mbox.\egroup
  }{2013}]{kwon2013prominent}
Kwon, S.; Cha, M.; Jung, K.; Chen, W.; and Wang, Y.
\newblock 2013.
\newblock Prominent features of rumor propagation in online social media.
\newblock In {\em Proceedings of ICDM},  1103--1108.

\bibitem[\protect\citeauthoryear{Ma \bgroup et al\mbox.\egroup
  }{2016}]{ma2016detecting}
Ma, J.; Gao, W.; Mitra, P.; Kwon, S.; Jansen, B.~J.; Wong, K.-F.; and Cha, M.
\newblock 2016.
\newblock Detecting rumors from microblogs with recurrent neural networks.
\newblock In {\em Proceedings of IJCAI},  3818--3824.

\bibitem[\protect\citeauthoryear{Maaten and
  Hinton}{2008}]{maaten2008visualizing}
Maaten, L. v.~d., and Hinton, G.
\newblock 2008.
\newblock Visualizing data using t-sne.
\newblock {\em Journal of machine learning research} 9(Nov):2579--2605.

\bibitem[\protect\citeauthoryear{Moore and Clayton}{2008}]{moore2008evaluating}
Moore, T., and Clayton, R.
\newblock 2008.
\newblock Evaluating the wisdom of crowds in assessing phishing websites.
\newblock In {\em Proceedings of FC},  16--30.

\bibitem[\protect\citeauthoryear{Ott \bgroup et al\mbox.\egroup
  }{2011}]{ott2011finding}
Ott, M.; Choi, Y.; Cardie, C.; and Hancock, J.~T.
\newblock 2011.
\newblock Finding deceptive opinion spam by any stretch of the imagination.
\newblock In {\em Proceedings of ACL},  309--319.

\bibitem[\protect\citeauthoryear{Pedregosa \bgroup et al\mbox.\egroup
  }{2011}]{scikit-learn}
Pedregosa, F.; Varoquaux, G.; Gramfort, A.; Michel, V.; Thirion, B.; Grisel,
  O.; Blondel, M.; Prettenhofer, P.; Weiss, R.; Dubourg, V.; Vanderplas, J.;
  Passos, A.; Cournapeau, D.; Brucher, M.; Perrot, M.; and Duchesnay, E.
\newblock 2011.
\newblock Scikit-learn: Machine learning in {P}ython.
\newblock {\em Journal of Machine Learning Research} 12:2825--2830.

\bibitem[\protect\citeauthoryear{Pennebaker \bgroup et al\mbox.\egroup
  }{2015}]{pennebaker2015development}
Pennebaker, J.~W.; Boyd, R.~L.; Jordan, K.; and Blackburn, K.
\newblock 2015.
\newblock The development and psychometric properties of liwc2015.
\newblock Technical report.

\bibitem[\protect\citeauthoryear{P{\'e}rez-Rosas \bgroup et al\mbox.\egroup
  }{2017}]{perez2017automatic}
P{\'e}rez-Rosas, V.; Kleinberg, B.; Lefevre, A.; and Mihalcea, R.
\newblock 2017.
\newblock Automatic detection of fake news.
\newblock {\em arXiv preprint arXiv:1708.07104}.

\bibitem[\protect\citeauthoryear{Popat \bgroup et al\mbox.\egroup
  }{2018}]{popat2018declare}
Popat, K.; Mukherjee, S.; Yates, A.; and Weikum, G.
\newblock 2018.
\newblock Declare: Debunking fake news and false claims using evidence-aware
  deep learning.
\newblock {\em arXiv preprint arXiv:1809.06416}.

\bibitem[\protect\citeauthoryear{Ruchansky, Seo, and
  Liu}{2017}]{ruchansky2017csi}
Ruchansky, N.; Seo, S.; and Liu, Y.
\newblock 2017.
\newblock Csi: A hybrid deep model for fake news detection.
\newblock In {\em Proceedings of CIKM},  797--806.

\bibitem[\protect\citeauthoryear{Shen \bgroup et al\mbox.\egroup
  }{2017}]{shen2017discovering}
Shen, H.; Ma, F.; Zhang, X.; Zong, L.; Liu, X.; and Liang, W.
\newblock 2017.
\newblock Discovering social spammers from multiple views.
\newblock {\em Neurocomputing} 225:49--57.

\bibitem[\protect\citeauthoryear{Shu \bgroup et al\mbox.\egroup
  }{2017}]{shu2017fake}
Shu, K.; Sliva, A.; Wang, S.; Tang, J.; and Liu, H.
\newblock 2017.
\newblock Fake news detection on social media: A data mining perspective.
\newblock {\em ACM SIGKDD Explorations Newsletter} 19(1):22--36.

\bibitem[\protect\citeauthoryear{Song \bgroup et al\mbox.\egroup
  }{2018}]{song2018directional}
Song, Y.; Shi, S.; Li, J.; and Zhang, H.
\newblock 2018.
\newblock Directional skip-gram: Explicitly distinguishing left and right
  context for word embeddings.
\newblock In {\em Proceedings of NAACL},  175--180.

\bibitem[\protect\citeauthoryear{Sutton and
  Barto}{1998}]{sutton1998introduction}
Sutton, R.~S., and Barto, A.~G.
\newblock 1998.
\newblock {\em Introduction to reinforcement learning}, volume 135.
\newblock MIT press Cambridge.

\bibitem[\protect\citeauthoryear{Sutton \bgroup et al\mbox.\egroup
  }{2000}]{sutton2000policy}
Sutton, R.~S.; McAllester, D.~A.; Singh, S.~P.; and Mansour, Y.
\newblock 2000.
\newblock Policy gradient methods for reinforcement learning with function
  approximation.
\newblock In {\em Proceedings of NIPS},  1057--1063.

\bibitem[\protect\citeauthoryear{Tacchini \bgroup et al\mbox.\egroup
  }{2017}]{tacchini2017some}
Tacchini, E.; Ballarin, G.; Della~Vedova, M.~L.; Moret, S.; and de~Alfaro, L.
\newblock 2017.
\newblock Some like it hoax: Automated fake news detection in social networks.
\newblock {\em arXiv preprint arXiv:1704.07506}.

\bibitem[\protect\citeauthoryear{Tschiatschek \bgroup et al\mbox.\egroup
  }{2018}]{tschiatschek2018fake}
Tschiatschek, S.; Singla, A.; Gomez~Rodriguez, M.; Merchant, A.; and Krause, A.
\newblock 2018.
\newblock Fake news detection in social networks via crowd signals.
\newblock In {\em Companion Proceedings of the The Web Conference},  517--524.

\bibitem[\protect\citeauthoryear{Wang \bgroup et al\mbox.\egroup
  }{2018}]{wang2018eann}
Wang, Y.; Ma, F.; Jin, Z.; Yuan, Y.; Xun, G.; Jha, K.; Su, L.; and Gao, J.
\newblock 2018.
\newblock Eann: Event adversarial neural networks for multi-modal fake news
  detection.
\newblock In {\em Proceedings of KDD},  849--857.

\bibitem[\protect\citeauthoryear{Wu, Li, and Wang}{2018}]{wu2018reinforced}
Wu, J.; Li, L.; and Wang, W.~Y.
\newblock 2018.
\newblock Reinforced co-training.
\newblock {\em arXiv preprint arXiv:1804.06035}.

\bibitem[\protect\citeauthoryear{Wu, Yang, and Zhu}{2015}]{wu2015false}
Wu, K.; Yang, S.; and Zhu, K.~Q.
\newblock 2015.
\newblock False rumors detection on sina weibo by propagation structures.
\newblock In {\em Proceedings of ICDE},  651--662.

\end{thebibliography}

\end{document}